\documentclass[journal=jctcce,manuscript=article,etalmode=truncate]{achemso}
\setkeys{acs}{etalmode=truncate,maxauthors=0,articletitle=true}
\setkeys{acs}{articletitle=true}

\usepackage{xcolor}
\usepackage{amsmath,amssymb}
\usepackage{caption}
\usepackage{subcaption}

\usepackage{algorithm2e}
\RestyleAlgo{ruled}

\newcommand{\dt}{\Delta t}
\newcommand{\Vk}{\mathbf{V}_\mathrm{krlv}}
\newcommand{\Hk}{\mathbf{H}_\mathrm{krlv}}

\newcommand{\bck}{\mat{c}_\mathrm{krlv}}
\newcommand{\TT}{\mathrm{T}}
\newcommand{\chebyp}{\Phi}

\usepackage[capitalize]{cleveref}
\crefname{figure}{Fig.}{Figs.}
\Crefname{figure}{Figure}{Figures}
\crefname{table}{Tab.}{Tabs.}
\Crefname{table}{Table}{Tables}
\crefname{equation}{Eq.}{Eqs.}
\Crefname{equation}{Equation}{Equations}
\crefname{section}{Sec.}{Secs.}
\Crefname{section}{Section}{Sections}
\crefname{paragraph}{Sec.}{Secs.}
\Crefname{paragraph}{Section}{Sections}
\crefname{algorithm}{Alg.}{Algs.}
\Crefname{algorithm}{Algorithm}{Algorithms}
\crefname{theorem}{Thm.}{Thms.}
\Crefname{theorem}{Theorem}{Theorems}
\crefname{corollary}{Cor.}{Cors.}
\Crefname{corollary}{Corollary}{Corollaries}

\newif\ifexporttikz
\newif\ifusetikz

\exporttikzfalse  
\usetikzfalse     

\ifexporttikz
    \usepackage{tikz}
    \usepackage{pgfplots}
    \pgfplotsset{compat=newest}
    \usepgfplotslibrary{external}
    \usepgfplotslibrary{fillbetween}
    \usetikzlibrary{patterns, intersections}
\else
    \ifusetikz
      \usepackage{tikzexternal}
      \tikzsetexternalprefix{fig-export/}
    \else
      \usepackage{graphicx}
    \fi
\fi

\ifusetikz
  
  \newcommand{\importlocalfigure}[1]{%
    \figname{#1}
    \input{fig-src/#1.tex}
  }
\else
  \newcommand{\importlocalfigure}[1]{%
    \includegraphics{fig-export/#1.pdf}
  }
\fi

\title{Approximate Exponential Integrators for Time-Dependent Equation-of-Motion Coupled Cluster Theory}

\author{David B. Williams-Young}
\affiliation{Applied Mathematics and Computational Research Division, Lawrence Berkeley National Laboratory, Berkeley, CA}
\email{dbwy@lbl.gov}

\author{Stephen H. Yuwono}
\affiliation{Department of Chemistry and Biochemistry, Florida State University, Tallahassee, FL}
\author{A. Eugene DePrince, III}
\affiliation{Department of Chemistry and Biochemistry, Florida State University, Tallahassee, FL}

\author{Chao Yang}
\affiliation{Applied Mathematics and Computational Research Division, Lawrence Berkeley National Laboratory, Berkeley, CA}

\newcommand{\op}[1]{\ensuremath \hat{#1}}
\newcommand{\stransop}[1]{\ensuremath \overline{#1}}
\newcommand{\mat}[1]{\ensuremath \boldsymbol{#1}}
\newcommand{\ket}[1]{\vert {#1}\rangle}
\newcommand{\bra}[1]{\langle {#1}\vert}

\renewcommand{\exp}[1]{\,\mathrm{exp}\!\left(#1\right)}
\newcommand{\eomH}[0]{\stransop{\mat H}_N}
\newcommand{\Nmol}[0]{N$_2$}

\begin{document}

\begin{abstract}

With growing demand for time-domain simulations of correlated many-body
systems, the development of efficient and stable integration schemes for the
time-dependent Schr\"odinger equation is of keen interest in modern electronic
structure theory. In the present work, we present two novel approaches for the
formation of the quantum propagator for time-dependent equation-of-motion
coupled cluster theory (TD-EOM-CC) based on the Chebyshev and Arnoldi
expansions of the complex, non-hermitian matrix exponential, respectively.  The
proposed algorithms are compared with the short-iterative Lanczos method of
Cooper, \emph{et al} [\emph{J. Phys. Chem. A} \textbf{2021} 125, 5438-5447],
the fourth-order Runge-Kutta method (RK4), and exact dynamics for a set of
small but challenging test problems. For each of the cases studied, both of the
proposed integration schemes demonstrate superior accuracy and efficiency relative to
the reference simulations.

\end{abstract}

\newpage
\section{Introduction}
\label{sec:intro}

In recent years, there has been renewed interest in the development of
efficient numerical methods to study the quantum dynamics of correlated
electrons in molecular and materials systems (see, e.g.,
Refs.~\citenum{Goings18_Real,Li20_Real} and references therein).  Under
particular approximations, it is possible to circumvent the direct solution of
the time-dependent Schr\"odinger equation (TDSE) in favor of time-dependent
perturbation theory (or ``frequency-domain" methods) which aims to implicitly
access quantum dynamics through probing the spectral structure of the
Hamiltonian operator. In the context of electronic structure theory, these
approaches include
linear-response,\cite{Dreuw05_Single,Olsen85_Linear,Datta95_Coupled}
polarization
propagator,\cite{Oddershede84_Polarization,Linderberg04_Propagators,Norman11_A}
and
equation-of-motion\cite{Ring04_The,Shavitt09_Many,Stanton93_The,Rico93_Single}
methods among others\cite{Trofimov06_ADC,Dreuw23_ADC,Peng21_GFCC}.  While these
methods can often be a powerful tool for the simulation and prediction of
observable phenomena such as spectroscopies, their veracity depends on the
applicability of their various approximations to accurately characterize
queried physical conditions.  
Further, the vast majority of these perturbative methods serve to access the
\emph{equilibrium} behaviour of electronic dynamics, leaving non-equilibrium
phenomena, such as charge migration\cite{Cederbaum99_Ultrafast}, inaccessible.
From a theoretical perspective, time-domain simulations do not suffer from
these deficiencies and may be straightforwardly extended to non-perturbative
and non-equilibrium regimes\cite{Goings18_Real,Li20_Real}.

Given the ability to faithfully represent physical conditions by a chosen
Hamiltonian, wave-function ansatz, and initial condition, the primary
challenges of time-domain electronic structure methods are practical rather
than theoretical. In contrast to frequency-domain methods which trade the
problem of temporal dynamics for the tools of numerical linear
algebra\cite{Coriani07_Linear,Kauczor11_On,Coriani12_Asymmetric,Kauczor13_Communication,VanBeeumen17_Model,Peng19_Approximate},
time-domain methods require explicit integration of the TDSE, which is
generally more resource intensive.  For hermitian discretizations of molecular
Hamiltonians, such as Hartree-Fock (real-time time-dependent HF, RT-TDHF
\cite{Micha94_Time,Li05_A}), density functional theory (RT-TDDFT)
\cite{Isborn07_Time}, and configuration interaction (TD-CI)
\cite{Krause05_Time,Schlegel07_Electronic,Lestrange18_Chapter,Sonk11_TDCI},
significant research effort has been afforded to the development of efficient
numerical methods to integrate the TDSE
\cite{Leforestier91_A,Gomez18_Propagators}.  In particular, approximate
exponential integrators based on polynomial
(Chebyshev\cite{TalEzer84_An,Leforestier91_A,WilliamsYoung16_Accelerating,Baer04_Real,Wang07_Density})
and Krylov subspace (short-iterative Lanczos\cite{Park86_Unitary}, SIL)
expansions of the quantum propagator are among the most widely used integration
techniques for hermitian quantum dynamics.  Exponential integrators are
powerful geometric techniques for the solution of linear ordinary differential
equations (ODE), such as the TDSE, as they preserve their exact
flow\cite{Hairer06_Geometric}, thereby allowing for much larger time-steps than
simpler, non-geometric integrators such as the fourth-order Runge-Kutta method
(RK4). In addition, these methods may also be formulated in such a way as to
only require knowledge of the action of a matrix-vector
product\cite{Leforestier91_A,Saad92_Analysis,AlMohy11_Computing,Hochbruck97_On},
thereby avoiding explicit materialization of the Hamiltonian matrix which is
generally large for correlated many-body wave-functions.

The situation is significantly more complex for non-hermitian Hamiltonian
discretizations such as those arising from coupled-cluster (CC) theory (see
Ref.~\citenum{Sverdrup23_Time} for a recent review).  Due to its simplicity and
low memory requirement, RK4 is generally the integrator of choice for
time-domain CC methods in the recent past\cite{Sverdrup23_Time}.
Symplectic\cite{Gray96_Symplectic,Park19_Equation,Pedersen19_Symplectic},
multistep\cite{Pathak23_RealTime}, and adaptive\cite{Wang22_Accelerating}
integrators for time-domain CC methods have been developed, and have yielded
significant efficiency improvements over their non-symplectic counterparts.
Exponential Runge-Kutta integrators have been explored in the context of
nonlinear time-dependent CC theory (TD-CC)\cite{Sato18_Time}, but have yet to
see wider adoption.  Recently, Cooper, \emph{et al.} \cite{Cooper21_Short}
suggested an approximate exponential integration scheme for time-dependent
equation-of-motion CC theory
(TD-EOM-CC)\cite{Sonk11_TDCI,Luppi12_Computation,Skeidsvoll22_Simulating,Sverdrup23_Time,Nascimento16_Linear,Nascimento19_A,Park19_Equation}
based on the hermitian SIL method to efficiently generate linear absorption
spectra for molecular systems.  Despite being only valid for hermitian
matrices, the proposed SIL approach was demonstrated to produce sufficiently
accurate spectra with relatively low subspace dimensions. However, the ability
of this scheme to produce faithful, long-time dynamics within TD-EOM-CC has not
been assessed, and is unlikely due to its hermitian ill-formation.  In this
work, we pursue the development and assessment of polynomial and non-hermitian
Krylov subspace (short-iterative \emph{Arnoldi}, SIA) methods for the complex
matrix exponential to enable the efficient and accurate simulation of
TD-EOM-CC.

The remainder of thie work is organized as follows. In \cref{sec:tdeomcc}, we
review the salient aspects of TD-EOM-CC theory relevant to the development of
efficient exponential integrators. In \cref{sec:exact_prop,sec:approx_prop} we
examine the properties of exact and approximate dynamics for the TD-EOM-CC ODE
and present the developmed integration schemes based on the Chebyshev
(\cref{sec:cheb}) and SIA (\cref{sec:sia}) expansions of the complex matrix
exponential. In \cref{sec:results}, we apply the developed integration schemes
to a set of small test problems and compare their verasity with exact dynamics
as well as previously employed SIL and RK4 methods. We conclude this work in
\cref{sec:conclusions} and offer outlook on future directions for approximate
exponential integrator development in TD-EOM-CC in the years to come.

\section{Theory and Methods}

\subsection{Time-Dependent Equation-of-Motion Coupled-Cluster Theory}
\label{sec:tdeomcc}

Time-dependent equation-of-motion coupled-cluster (TD-EOM-CC) theory is a
general time-domain reformulation of many-body quantum mechanics capable of
simulating the dynamics of both
time-dependent\cite{Sonk11_TDCI,Luppi12_Computation,Skeidsvoll22_Simulating,Sverdrup23_Time}
and time-independent\cite{Nascimento16_Linear,Nascimento19_A} Hamiltonians.  In
this work, we consider the moment-based formulation\cite{Nascimento16_Linear}
of TD-EOM-CC to compute the spectral function,
\begin{equation}
f(\omega) = \frac{2}{3}\omega\,\int_{-\infty}^\infty \mathrm{d}t\, e^{-i\omega t} S(t), \label{eq:osc_str_func}
\end{equation}
where $S(t) = \langle \tilde M(0) \vert M(-t)\rangle = \langle \tilde M(t) \vert M(0)\rangle$
is the autocorrelation function. Here, $\ket{M(t)}$
($\bra{\tilde{M}(t)}$) is (the dual of) the time-dependent moment function
which describes the propagation of weak perturbations throughout the many-body
system. We note for clarity that, due to the nonhermiticity of the CC formalism, $\bra{\tilde{M}(t)}$
is not the complex conjugate of $\ket{M(t)}$. Additionally, throughout this paper, we chose
$S(t)$ to be $\langle \tilde M(0) \vert M(-t)\rangle$, although $\langle \tilde M(t) | M(0)\rangle$
is also valid.
$\ket{M(t)}$ ($\bra{\tilde{M}(t)}$) may generally be described via a linear expansion of
(de-)excitations from a reference state $\ket{0}$ (typically taken to be HF),
\begin{align}
\ket{M(t)} &= \left( 
  m_0(t) + 
  \sum_{ai} m_i^a(t) c_a^\dagger c_i + 
  \frac{1}{4}\sum_{abij} m_{ij}^{ab}(t) c_a^\dagger c_b^\dagger c_j c_i + 
  \cdots
\right)\ket{0} \label{eq:moment_amplitudes}\\
\bra{\tilde{M}(t)} &= \bra{0}\left(
  \tilde{m}_0(t) + 
  \sum_{ai}\tilde{m}_a^i(t) c_i^\dagger c_a  + 
  \frac{1}{4}\sum_{abij}\tilde{m}_{ab}^{ij}(t) c_i^\dagger c_j^\dagger c_b c_a + 
  \cdots 
\right)
\end{align} 
where $m_0$ ($\tilde m_0$), $m_i^a$ ($\tilde m_a^i$) and $m_{ij}^{ab}$ ($\tilde
m_{ab}^{ij}$) are time-dependent (de-)excitation amplitudes, $c_p$
($c_p^\dagger$) is the fermionic  annihilation (creation) operator associated
with the spin-orbital $p$, and the indices $i,j,\ldots$ and $a,b.\ldots$ denote occupied
and virtual spin-orbitals relative to $\ket{0}$.  In this work, we truncate
\cref{eq:moment_amplitudes} to only include up to double excitations from the
reference, resulting in the TD-EOM-CCSD approach.

Within the TD-EOM-CC formalism, the moment excitation and de-excitation amplitudes obey the following set of coupled,
linear-time-invariant (LTI) ODEs\cite{Nascimento16_Linear}
\begin{equation}
\partial_t \mat{m}(t) = -i\eomH\mat{m}(t), \quad \mat{m}(t) = 
  \begin{bmatrix} 
  m_0(t) \\ 
  \{m_i^a(t)\} \\ 
  \{m_{ij}^{ab}(t)\} 
  \end{bmatrix} \in\mathbb{C}^n,
\label{eq:td_eom_coeffs}
\end{equation}
and their left-hand counterparts
\begin{equation}
\partial_t \tilde{\mat{m}}(t) = i\eomH^\TT \tilde{\mat{m}}(t), \quad \tilde{\mat{m}}(t) = 
  \begin{bmatrix} 
  \tilde m_0(t) \\ 
  \{\tilde m_i^a(t)\} \\ 
  \{\tilde m_{ij}^{ab}(t)\} 
  \end{bmatrix} \in\mathbb{C}^n,
\label{eq:td_eom_coeffs_tilde}
\end{equation}
where $\eomH\in\mathbb{C}^{n\times n}$ is the non-hermitian, normal-ordered,
similarity-transformed Hamiltonian represented in the basis of Slater
determinants \cite{Shavitt09_Many,Stanton93_The}.  From the moment
state-vectors,
$\mat m(t)$ and $\tilde{\mat{m}}(t)$,
$S(t)$ of \cref{eq:osc_str_func} may be evaluated as
\begin{equation}
  S(t) = \tilde{\mat m}^\TT \mat m(-t), \label{eq:property_eval}
\end{equation}
where we have taken $\tilde{\mat m}\equiv \tilde{\mat m}(0)$.
It is worth mentioning that the TD-EOM-CCSD formalism used here requires
propagating only the right- or left-hand moment amplitudes [in this case, the
right-hand amplitudes following \cref{eq:td_eom_coeffs}].
While \cref{eq:osc_str_func} is perturbatively derived from Fermi's Golden
Rule\cite{Nascimento16_Linear}, time evolution of $\ket{M(t)}$ via
\cref{eq:td_eom_coeffs} also serves as a useful model for the development of
both LTI and non-LTI integration techniques for TD-EOM-CC methods as it
formally consists of the same algorithmic components that are required for the
simulation of time-dependent Hamiltonians
\cite{Sonk11_TDCI,Luppi12_Computation,Skeidsvoll22_Simulating,Sverdrup23_Time}.

When specified as an initial value problem, \cref{eq:td_eom_coeffs} admits an
analytic solution
\begin{equation}
\mat{m}(t) = \exp{-i\eomH t}\mat{m}(0) \label{eq:td_coeffs_sol}
\end{equation} 
where $\exp{-i\eomH t}$ is the quantum propagator and $\mathrm{exp}$ is the
matrix exponential defined in the canonical way\cite{Moler03_Nineteen}.  We
refer the reader to Refs.\citenum{Nascimento16_Linear,Nascimento19_A} for
discussions pertaining to the choices of initial conditions for
\cref{eq:td_coeffs_sol} to simulate various spectroscopic properties. In this
work, we consider the dipole initial conditions\cite{Nascimento16_Linear}
induced by
\begin{equation}
  \ket{M(0)} = \stransop{\mu}\ket{0}, \quad
  \bra{\tilde{M}(0)} = \bra{0}(1 + \op\Lambda)\stransop{\mu}, \quad
  \stransop{\mu} = \exp{-\op T} \op \mu \exp{\op T},
\end{equation}
where $\op T$ and $\op \Lambda$ are the ground-state CC excitation and
de-excitation operators (again truncated at double excitation/de-excitations in
this work), and $\op \mu$ is a particular component of the electronic dipole
operator.

\subsection{Exact Matrix Exponential}
\label{sec:exact_prop}

When $\eomH$ is small enough to be formed explicitly in memory,
\cref{eq:td_coeffs_sol} may be directly evaluated as
\begin{equation}
  \mat{m}^\mathrm{ex}(t) = \mat R \exp{-i\mat\Omega t} \mat w^\mathrm{ex}, \quad 
  \mat w^\mathrm{ex} = \mat L \mat m(0)\label{eq:exact_prop}
\end{equation}
where $\mat \Omega\in\mathbb{C}^{n\times n}$ is the diagonal matrix of EOM-CC
eigenvalues, $\Omega = \{\omega_I \in \mathbb{C}\}_{I=1}^n$, and $\mat L,\mat
R\in\mathbb{C}^{n\times n}$ are the full, biorthogonal set of corresponding
left and right eigenvectors safisfying the equations
\cite{Stanton93_The,Shavitt09_Many}
\begin{equation}
  \eomH \mat R = \mat R \mat \Omega,\quad
  \mat L \eomH = \mat \Omega \mat L,\quad
  \mat L \mat R = \mat I
\end{equation}
where $\mat I\in\mathbb{C}^{n\times n}$ is the identity-matrix. As $\mat\Omega$
is a diagonal matrix, $\exp{-i\mat\Omega t}$ is simply the diagonal matrix with
entries $e^{-i\omega_I t}$. Insertion of \cref{eq:exact_prop} into
\cref{eq:property_eval} yields the following simple expression for the exact
autocorrelation function
\begin{equation}
  S_\mathrm{ex}(t) = \tilde{\mat{w}}^\mathrm{ex,T} \exp{i\mat\Omega t} \mat w^\mathrm{ex}, \quad
  \tilde{\mat w}^\mathrm{ex} = \mat R^\TT \tilde{\mat m}.
  \label{eq:exact_s}
\end{equation}

As a non-hermitian matrix, $\eomH$ is not guaranteed to have real eigenvalues
if the many-electron basis is truncated, and as such, \cref{eq:exact_prop} (and
by extension \cref{eq:td_coeffs_sol}) is not guaranteed to be unitary
(norm-preserving) and will generally yield dissipative or divergent dynamics
along EOM-CC modes with $\Im\omega_I\neq 0$ (see, \emph{e.g.}, a recent study
in Ref. \citenum{DePrince23_Complex}). However, it has been shown that,
\cite{Kjonstad17_Crossing,Thomas21_Complex} barring suboptimal ground-state CC
solutions or the presence of conical intersections, $\eomH$ typically admits a
real spectrum representing physical excited states and thus,
\cref{eq:exact_prop} is unitary in exact arithmetic.

\subsection{Approximate Exponential Integrators}
\label{sec:approx_prop}

While \cref{eq:exact_prop} is an exact solution to the LTI TD-EOM-CC dynamics
considered in this work, it requires the full diagonalization of $\eomH$. As
the memory requirement associated with the EOM-CCSD $\eomH$ grows $O(N^8)$ with
system size, full diagonalization is impractical for all but the smallest
problems.  For some systems, it is possible to integrate the TD-EOM-CC
equations in a subspace spanned by a small number of states such that full
diagonalization is not
required.\cite{Sonk11_TDCI,Luppi12_Computation,Skeidsvoll22_Simulating,Sverdrup23_Time}
However, if a large number of states are required or spectral regions of
interest are densely populated or spectrally interior, this approach also
becomes impractical.

Matrix exponentiation is a challenging numerical linear algebra problem, and
the past half century has yielded a wealth of research into the development of
efficient
implicit\cite{Leforestier91_A,Saad92_Analysis,AlMohy11_Computing,Hochbruck97_On}
and direct\cite{Moler03_Nineteen} methods both for hermitian and non-hermitian
matrices.  In this work, we will consider subspace approaches for evaluation of
the complex, non-hermitian matrix exponential generally taking the form
\begin{equation}
  \mat m(t+\delta t) = \exp{-i\eomH \delta t}\mat{m}(t) \approx \mat V \mat c(\delta t), \label{eq:subspace_exp}
\end{equation}
where $\mat V\in\mathbb{C}^{n\times k}$ is a $k$-dimensional subspace (with
$k\ll n$) generated by the action of $-i\eomH$ onto the current state
vector, $\mat m(t)$, and $\mat c(\delta t)\in\mathbb{C}^k$ is a time-varying
coefficient vector.  Given the ability to implicitly form $\mat \sigma
\leftarrow \eomH\mat v$ (i.e. a ``$\sigma$ build"), which is a standard
algorithmic component of any EOM-CC
implementation\cite{Stanton93_The,Shavitt09_Many}, the implementations of
\cref{eq:subspace_exp} considered in this work will not require materialization
of $\eomH$ in memory. Within the subspace ansatz, \cref{eq:property_eval}
becomes 
\begin{equation}
  S(t+\delta t) \approx \tilde{\mat w}^T \mat c(-\delta t), \quad \tilde{\mat w} = \mat{V}^T \tilde{\mat m}\in\mathbb{C}^k, \label{eq:approx_property}
\end{equation}
where $\tilde{\mat w}$ is time-independent for fixed $\mat V$.

For a particular expansion order $k$ and state vector $\mat m(t)$,
\cref{eq:subspace_exp} will generally be valid for $\vert\delta t\vert \leq
\vert\Delta t\vert$, where $\Delta t$ will be referred to as a \emph{macro
time-step} in the following.  Within this prescription, the total simulation
length, $\mathcal{T}$, will be partitioned into subintervals
$\{\mathcal{T}_i=[t_{i},t_{i+1}]\}$ where $t_0=0$, $t_i=t_{i-1} + \Delta t_i$,
and $\Delta t_i$ is the macro time step for the $i$-th interval.  The
relationship between $k$ and $\Delta t$ is method-dependent, and will be
discussed for both the Chebyshev and Arnoldi integrators below. Due to the
factorization of the time-dependence into $\mat{c}(t)$, a general property of
truncated expansions such as \cref{eq:subspace_exp} is in their ability to
interpolate within each $\mathcal{T}_i$ without requiring additional $\sigma$
builds\cite{Leforestier91_A}. This property is particularly advantageous for
methods such as EOM-CCSD in which the computational complexity of $\sigma$
formation scales $O(N^6)$ with system size\cite{Stanton93_The,Shavitt09_Many}.
For each $\mathcal{T}_i$, a single $\mat V$ is computed and the propagator may
be interpolated to arbitrary temporal resolution by varying the corresponding
coefficients. For each of the intermediate time intervals ($i>0$), the
approximation of $\mat m(t_{i+1})$ generated from the endpoint of
$\mathcal{T}_i$ is used as the starting vector to generate $\mat V$ for
$\mathcal{T}_{i+1}$.

\subsubsection{Chebyshev Time Integration}
\label{sec:cheb}

The use of the Chebyshev expansion to evaluate the quantum propagator for
hermitian Hamiltonians is well established and is among the most efficient
known strategies for integrating LTI variants of the
TDSE\cite{TalEzer84_An,Leforestier91_A,WilliamsYoung16_Accelerating,Baer04_Real,Wang07_Density}.
In this work, we demonstrate that this approach is also applicable to
non-hermitian Hamiltonians with real spectra.  Chebyshev polynomials of the
first kind, $\{\chebyp_p\}$, given by the recurrence 
\begin{equation}
  \chebyp_0(z) = 1, \ \ 
  \chebyp_1(z) = z, \ \ 
  \chebyp_{p+1}(z)=2z\chebyp_p(z) - \chebyp_{p-1}(z),
  \label{eq:cheby3term}
\end{equation}
are a powerful tool in the approximation of scalar and matrix functions on the
real-line as they form the unique approximation basis which minimizes the
uniform (infinity) norm on $[-1,1]$ at a particular
order\cite{Burden15_Numerical}. 
In the Chebyshev basis, the TD-EOM-CC propagator acting on a general vector
$\mat v$ may be exactly expanded as\cite{TalEzer84_An,Leforestier91_A}
\begin{equation}
  \exp{-i\eomH \delta t}\mat v = 
    e^{-i \gamma_+ \delta t} 
      \sum_{p=0}^\infty (2 - \delta_{p0}) J_p(\gamma_- \delta t) 
      \chebyp_p(-i\tilde{\mat{H}}_N) \mat v \label{eq:chebyprop}
\end{equation}
where $\gamma_\pm = \frac{1}{2}(\omega_\mathrm{max} \pm \omega_\mathrm{min})$,
$\omega_\mathrm{min/max}$ are the minimum/maximum eigenvalues of $\eomH$,
$\delta_{k0}$ is a Kronecker delta, $J_p$ is the $p$-th Bessel function of the
first kind, and $\tilde{\mat{H}}_N = \gamma_-^{-1}(\eomH - \gamma_+\mat I)$ is
an auxilary matrix that scales the spectrum of $\eomH$ from
$[\omega_\mathrm{min},\omega_\mathrm{max}]\rightarrow[-1,1]$ such that the
image of $\chebyp_p$ remains on the unit disk.  Practically, $\tilde{\mat{H}}_N$ need
not be formed explicitly (see \cref{alg:chebyexp}) and $\gamma_\pm$ need not be
computed from exact eigenvalues and can be approximated using standard
techniques
\cite{Sorensen97_Implicitly,Lehoucq98_ARPACK,Kjonstad20_Accelerated,Zuev15_New,Caricato10_A}
as long as the mapped spectral bounds are contained in $[-1,1]$. 

\begin{figure}[t]
  \centering
  \begin{minipage}{0.48\textwidth}
    \centering
    \importlocalfigure{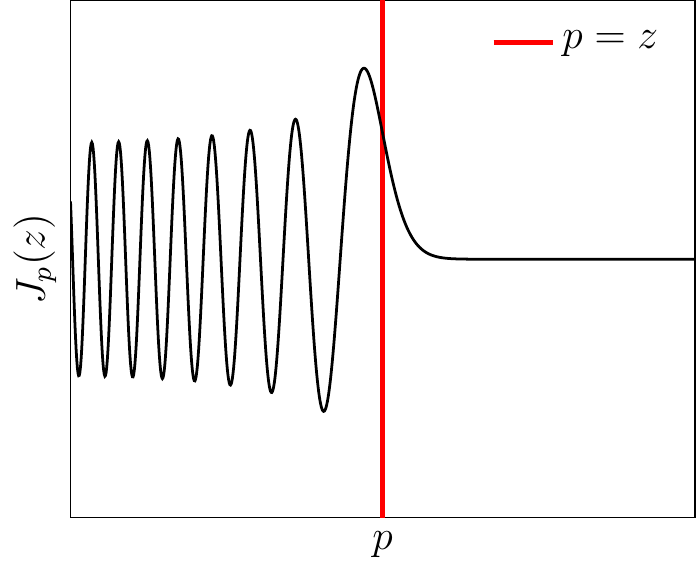}
  \end{minipage}
  \caption{Graphical depiction of the order decay behaviour of Bessel functions
  of the first kind for fixed argument. The function is highly oscillatory for
  $p<z$ but decays exponentially for $p>z$.}
  \label{fig:bessel_decay}
\end{figure}

In practice, the sum in \cref{eq:chebyprop} is truncated to a finite order $k$,
yielding a compact representation of the propagator in the Chebyshev basis,
$\mat{V}_\mathrm{cheb}=[\mat v^\mathrm{cheb}_0,\mat v^\mathrm{cheb}_1, \mat
v^\mathrm{cheb}_2, \cdots, \mat v^\mathrm{cheb}_{k-1}]$, given by
\begin{equation}
  \mat v_p^\mathrm{cheb} = \chebyp_p(-i\tilde{\mat H}_N)\mat m(t), \quad
  c_p^\mathrm{cheb}(\delta t) = e^{-i\gamma_+ \delta t}(2-\delta_{p0}) J_p(\gamma_- \delta t). \label{eq:cheby_basis_expand}
\end{equation}
The truncation error at the interval endpoint ($t + \Delta t$) of the Chebyshev
expansion can be shown\cite{BADER2022,lubich2008} to be bounded by
\begin{equation}
  C(\Delta t) = 2\|\mat v\|\sum_{p=k}^{\infty}
  \vert J_p(\gamma_-\Delta t)\vert. \label{eq:chebybnd_bessel}
\end{equation}
For fixed argument, $J_p(z)$ is highly oscillatory for $p< z$ but decays
exponentially for $p > z$, as depicted in \cref{fig:bessel_decay}.  We note
that for even (odd) $p$, $J_p$ is an even (odd) function about zero.
Therefore, for $p$ sufficiently larger than $\vert\gamma_- \Delta t\vert$, we
may approximate $C(\dt)\approx 2\|\mat v\| \vert J_p(\gamma_- \Delta t)\vert$.
Given a desired step size, $\Delta t_\mathrm{cheb}$, and error threshold
$\varepsilon^\mathrm{cheb}$, we may use this approximation to select
$k>\vert\gamma_- \Delta t_\mathrm{cheb}\vert$ such that $\vert J_{k}(\gamma_-
\Delta t_\mathrm{cheb})\vert<\frac{\varepsilon^\mathrm{cheb}}{2\|\mat v\|}$. 

\begin{algorithm}[t]
\caption{Evaluation of \cref{eq:approx_property,eq:chebyprop} via the Chebyshev Expansion}
\label{alg:chebyexp}
\SetKwInput{KwData}{Input}
\SetKwInput{KwResult}{Returns}
\KwData{%
  $\eomH\in\mathbb{C}^{n\times n}$, %
  $\gamma_\pm\in\mathbb{R}$, %
  $\Delta t_\mathrm{cheb}\in\mathbb{R}$, %
  Truncation order $k\in\mathbb{Z}^+$, %
  $\tilde{\mat m}\in\mathbb{C}^n$, %
  and current state vector $\mat m(t)\in\mathbb{C}^n$%
}
\KwResult{$\tilde{\mat w}^\mathrm{cheb}$, $\mat m_*\approx \mat m(t+\Delta t_\mathrm{cheb})$} 
~\\
  \nl $\alpha\leftarrow \gamma_-\Delta t_\mathrm{cheb}$\\
  \nl $\mat v_- \leftarrow \mat m (t)$ \\
  \nl $\tilde{w}^\mathrm{cheb}_0 \leftarrow \mat v_-^\TT \tilde{\mat m}$ \\
  \nl $\mat m_* \leftarrow J_0(\alpha)\mat v_-$ \\
  \nl $\mat \sigma \leftarrow \eomH \mat v_-$ \\
  \nl $\mat v_0 \leftarrow -i\gamma_-^{-1}(\mat \sigma - \gamma_+ \mat v_-)$ \\
  \nl $\tilde{w}^\mathrm{cheb}_1 \leftarrow \mat v_0^\TT \tilde{\mat m}$ \\
  \nl $\mat m_* \leftarrow \mat m_* + 2J_1(\alpha)\mat v_0$ \\
  \For{$p\in[2,k)$}{
    \nl $\mat \sigma \leftarrow \eomH \mat v_0$ \\
    \nl $\mat v_+ \leftarrow -2i\gamma_-^{-1}(\mat \sigma - \gamma_+ \mat v_0) + \mat v_-$\\
    \nl $\tilde{w}^\mathrm{cheb}_p \leftarrow \mat v_+^\TT \tilde{\mat m}$ \\
    \nl $\mat m_* \leftarrow \mat m_* + 2J_p(\alpha)\mat v_+$ \\
    \nl $\mat v_- \leftarrow \mat v_0$ \\
    \nl $\mat v_0 \leftarrow \mat v_+$ \\
  }
  \nl $\mat m_* \leftarrow e^{-\gamma_+\Delta t_\mathrm{cheb}}\mat m_*$
\end{algorithm}

As $\Delta t_\mathrm{cheb}$ is fixed, $\mathcal{T}$ may be evenly partitioned
into $\lceil \frac{\mathcal T}{\vert\Delta t_\mathrm{cheb}\vert} \rceil$
intervals. The Chebyshev subspace vectors may be efficiently evaluated using
only $k$ $\sigma$-builds (\cref{alg:chebyexp}), thus the total $\sigma$-build
cost for this method is $\lceil \frac{\mathcal T}{\vert\Delta
t_\mathrm{cheb}\vert} \rceil\cdot k$.  Another important aspect of the
Chebyshev method is that, due to fact that the expressions in
\cref{eq:cheby_basis_expand} are analytic, one need not materialize $\mat
V_\mathrm{cheb}$ in memory.  Instead, one may evaluate $\tilde{\mat
w}^\mathrm{cheb} = \mat V_\mathrm{cheb}^\TT \tilde{\mat m}$
(\cref{eq:approx_property}) directly as the subspace is generated, as is shown
in \cref{alg:chebyexp}, thus changing the memory requirement from $O(kn)$ to
$O(3n)$. As it is often the case that one requires high-order Chebyshev
polynomials ($\gg 3$) to accurately approximate the matrix exponential, this realization leads to
a drastic reduction in memory consumption for large systems.

\subsubsection{Short Iterative Arnoldi Time Integration}
\label{sec:sia}

Considering the spectral decomposition of the exact propagator given in
\cref{sec:exact_prop}, it is expected that the Chebyshev method discussed in
\cref{sec:cheb} will be most effective when $\Omega$ is nearly uniformly
distributed within $[\omega_\mathrm{min},\omega_\mathrm{max}]$, due to the fact
that the Chebyshev basis minimizes the uniform function norm. If $\Omega$ is
clustered, Krylov subspace techniques for the formation of the exponential
propagator are often more effective\cite{Saad92_Analysis}. The basic principle
behind Krylov approximation techniques for matrix-functions is rooted in the
generation of a $k$-dimensional, orthonormal basis,
$\Vk=[\mat{v}_0^\mathrm{krlv},\mat{v}_1^\mathrm{krlv},\cdots,\mat{v}_{k-1}^\mathrm{krlv}]$,
for the Krylov subspace
\begin{equation}
  \mathcal{K}^k(\eomH,\mat v_0) = 
     \{\mat v_0,\eomH \mat v_0,\eomH^2 \mat v_0,...,\eomH^{k-1}\mat v_0\}.
  \label{eq:krylov}
\end{equation}
where $\mat v_0\in\mathbb{C}^n$ is an arbitrary vector with $\|\mat v_0\|=1$.
Given $\Vk$, one may form a subspace-projected Hamiltonian,
\begin{equation}
  \Hk = \Vk^\dagger \eomH \Vk \in\mathbb{C}^{k\times k}
\end{equation}
and approximate the action of the matrix exponential as\cite{Saad92_Analysis}
\begin{equation}
  \exp{-i\eomH \delta t} \mat v \approx \Vk \bck(\delta t), \quad
  \bck(\delta t) = \|v\| \exp{-i\Hk \delta t} e_1,
  \label{eq:fAapprox}
\end{equation}
where $e_1$ is the first column of a $k\times k$ identity matrix and $\Vk$ is
the Krylov subspace generated from $\mat v_0 = \mat v/\|\mat v\|$.  Given that
$k\ll n$, the exponential in \cref{eq:fAapprox} may be efficiently evaluated
via \cref{eq:exact_prop}.

For hermitian matrices, $\Vk$ can be efficiently generated by the Lanczos
iteration\cite{Stewart01_Matrix}, $\Hk$ is a tridiagonal matrix, and both $\Hk$
and $\Vk$ may be formed implicitly via a simple three-term recursion. For the
approximation of the propagator, this approach has come be known as the
short-iterative Lanczos (SIL) method\cite{Park86_Unitary}. Here, we present an
analogous scheme for the exponential propagator based on the Arnoldi
iteration\cite{Stewart01_Matrix,Saad11_Numerical}, which is a general Krylov
subspace technique which extends to both hermitian and non-hermitian matrices.
We will refer to this approach as the short-iterative Arnoldi (SIA) method in
the following. Instead of a tridiagonal matrix, the Arnoldi method produces an
upper Hessenburg matrix via the recursion
\begin{equation}
  \eomH \Vk = \Vk \Hk + \beta_{k+1} \mat v^\mathrm{krlv}_{k+1}e_k^\TT
\end{equation}
where $e_k$ is the $k$-th column of the $k\times k$ identity matrix and
$\beta_{k+1}\mat v^\mathrm{krlv}_{k+1}$ is the residual
\begin{equation}
  \beta_{k+1}\mat v^\mathrm{krlv}_{k+1} = (I-\Vk\Vk^\dagger)\eomH \mat v^\mathrm{krlv}_k,
\end{equation}
with $\|\mat v^\mathrm{krlv}_{k+1}\|=1$. If $\eomH$ were an hermitian matrix,
$\Hk$ would be tridiagonal and $\Vk$ would span the same subspace as the one
produced by the Lanczos iteration in exact arithmetic. 

\begin{algorithm}[t]
\caption{The Arnoldi Iteration}
\label{alg:arnoldi}
\SetKwInput{KwData}{Input}
\SetKwInput{KwResult}{Returns}
\KwData{%
  $\eomH\in\mathbb{C}^{n\times n}$,  %
  $\mat v_0\in\mathbb{C}^n$ with $\|\mat v_0\|_2=1$, %
  Krylov dimension $k\in\mathbb{Z}^+$%
}
\KwResult{%
  Krylov basis $\mat V_k\in\mathbb{C}^{n\times k}$, %
  Projected Hamiltonian $\Hk\in\mathbb{C}^{k\times k}$.%
}
~\\
  \nl $\mat v_0^\mathrm{krlv} \leftarrow \mat v_0$ \\
  \nl $\Vk\leftarrow[\mat v^\mathrm{krlv}_0]$ \\
  \For{$p\in[0,k-1)$} {
    \nl $\mat \sigma \leftarrow \eomH \mat v^\mathrm{krlv}_{p}$ \\
    \nl $\mat h_1\leftarrow \Vk^\TT\mat \sigma$ \tcp*{Classical Gram-Schmidt} 
    \nl $\mat \sigma\leftarrow \mat \sigma - \Vk\mat h_1$ \\
    \nl $\mat h_2\leftarrow \Vk^\TT\mat \sigma$ \tcp*{Reorthogonalization} 
    \nl $\mat \sigma\leftarrow \mat \sigma - \Vk\mat h_2$ \\
    \nl $\beta \leftarrow \|\mat \sigma\|_2$  \\
    \nl $\Hk(0:p, p) \leftarrow \mat h_1 + \mat h_2$ \\
    \nl $\Hk(p+1, p)\leftarrow \beta$ \\
    \nl $\mat v_{p+1}^\mathrm{krlv} \leftarrow \beta^{-1}\mat \sigma$ \\
    \nl $\Vk\leftarrow[\Vk, \mat v^\mathrm{krlv}_{p+1}]$ \\
  }
\end{algorithm}

Much like the Lanczos iteration, $\Hk$ may also be formed incrementally via the
Arnoldi iteration as shown in \cref{alg:arnoldi}. However, unlike the 3-term
recurrence used in the Lanczos method, the Arnoldi iteration requires
\emph{explicit} orthogonalization of newly produced subspace vectors as opposed
to the implicit orthgonalization generated by Lanczos. As the Arnoldi method is
guaranteed to produce orthonormal basis via explicit orthogonalization, it is
often more numerically stable even for hermitian
problems~\cite{paige1976,selectorth,SIMON1984101}.  In this work, we have
utilized the classical Gram-Schmidt method with reorthogonaliztation to perform
the explicit basis orthogonalization~\cite{dgks}.  There exist
non-hermitian extensions of the Lanczos method \cite{Saad82_The} which produce
simultaneous, biorthogonal approximations for the left- and right-hand
eigenspaces of non-hermitian matrices and have seen successful applications in
both frequency domain CC applications\cite{Coriani12_Asymmetric} as well as in
state selection for TD-EOM-CC\cite{Skeidsvoll22_Simulating}. However, the
biorthogonalization requirements of these methods can often be numerically
unstable~\cite{LanczosLookahead,gutknecht92,VANDERVEEN1995605}, and as such, we
expect the Arnoldi method to yield superior numerical stability in finite
precision~\cite{arioli96}.

It has been shown~\cite{Saad92_Analysis} that the error produced by
\cref{eq:fAapprox} can be bounded by the right hand side of the following
inequality
\begin{equation}
  \| \exp{i\eomH\delta t} \mat v_0 - \Vk \exp{i\Hk\delta t} e_1 \|_2 \leq 
  2 \beta_{k+1}  (\delta t \rho)^k \max (1,e^{\mu(-\eomH)\delta t}),
  \label{eq:Arnbound}
\end{equation}
where $\mu(\eomH)$ is the largest eigenvalue of $(\eomH + \eomH^\dagger)/2$ and
$\rho = \|\eomH\|_2$. Although tighter bounds can be
found\cite{Hochbruck97_On}, the bound given in \eqref{eq:Arnbound} is more
instructive. It shows that the approximation error made in an Arnold time
integrator depends on the departure of $\Vk$ from an invariant subspace of
$\eomH$, which is measured by $\beta_{k+1}$, the step size or time window
$\delta t$ as well as the spectral radius of $\eomH$, measured by $\rho$ and
$\mu(\eomH)$.

Unlike the Chebyshev method, where the expansion coefficients are known ahead
of time, the coefficients for SIA are related to the spectrum of $\Hk$, which
itself is dependent on $\mat v$ (the current state vector, $\mat m(t)$, in the
context of \cref{eq:subspace_exp}). As such, it is canonical to adopt a dynamic
time-stepping approach where the Krylov subspace dimenion ($k$) is fixed before
the simulation and each $\Delta t_i$ corresponding to $\mathcal{T}_i$ is
determined dynamically throughout the time propagation. As \cref{eq:Arnbound}
is only a loose bound, its practical ability to determine $\Delta t$ is
limited.  Given that the Arnoldi method produces successively more accurate
Krylov subspaces with increasing $k$, a more practical error bound is given by
$c^\mathrm{krlv}_k(\Delta t)$, which measures the potential for projections of
the exact matrix-exponential onto vectors outside the Krylov subspace.
Therefore, as has been successfully applied to the SIL
method\cite{Cooper21_Short}, a reasonable choice for the step size is the
largest $\Delta t$ such that $\vert c^\mathrm{krlv}_k(\Delta
t)\vert<\varepsilon^\mathrm{krylov}$, where
$\varepsilon^\mathrm{krylov}\in\mathbb{R}^+$ is a chosen error threshold.

Another side effect of the non-analytic nature of the SIA coefficients is that,
unlike $\mathbf{V}_\mathrm{cheb}$, $\Vk$ must be materialized in memory and
\cref{eq:approx_property,eq:chebyprop} must be evaluated explicitly. As such,
the memory requirement assococaited with SIA will grow $O(kn)$ with basis
dimension.  However, as will be demonstrated in \cref{sec:results}, the SIA
method will generally require fewer $\sigma$ builds than the Chebyshev method
to achieve commensurate integration accuracy. 

\section{Results}
\label{sec:results}

To assess the efficacy of the Chebyshev and SIA TD-EOM-CC integrators developed
in this work, we compare the accuracy and efficiency of these methods for two
test systems, {\Nmol~(1.1 \AA)} and MgF (1.6 \AA), relative to exact dynamics
(\cref{eq:exact_prop}) as well as RK4 and the TD-EOM-CC SIL method of
Ref.~\citenum{Cooper21_Short}.  Each of these systems were treated at the
EOM-CCSD level of theory with the minimum STO-3G basis
set\cite{Pople69_2657,Pople70_2769} to allow for practical comparisons with
exact dynamics.  All ground-state CC calculations were performed using a
prototype Python implementation interfaced with the HF and integral
transformation routines in the \textsc{Psi4} software package
\cite{Sherrill20_184108} and geometries were aligned along the $z$-cartesian
axis without the use of point-group symmetry.  At their respective geometries,
both of these systems exhibit real-valued EOM-CC spectra.  All simulations in
this work were performed using $\varepsilon^\mathrm{cheb}=10^{-16}$ and
$\varepsilon^\mathrm{krylov}=10^{-6}$ (for both SIL and SIA) for a duration of
$\mathcal{T}=1350$ $E_h^{-1}$ ($\approx$ 32 fs).

\begin{figure}[t]
  \centering
  \begin{subfigure}[b]{0.48\textwidth}
    \centering
    \importlocalfigure{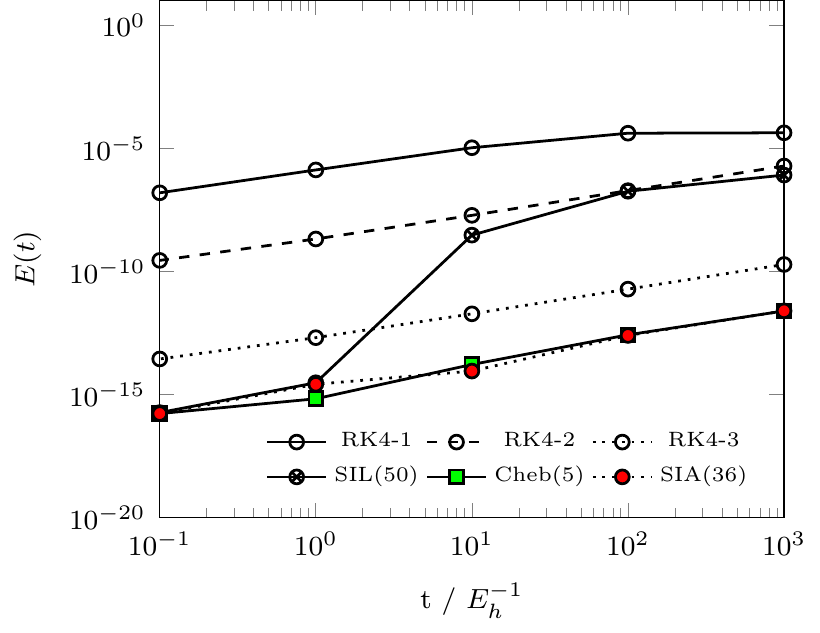}
    \caption{\Nmol}
  \end{subfigure}\hfill%
  \begin{subfigure}[b]{0.48\textwidth}
    \centering
    \importlocalfigure{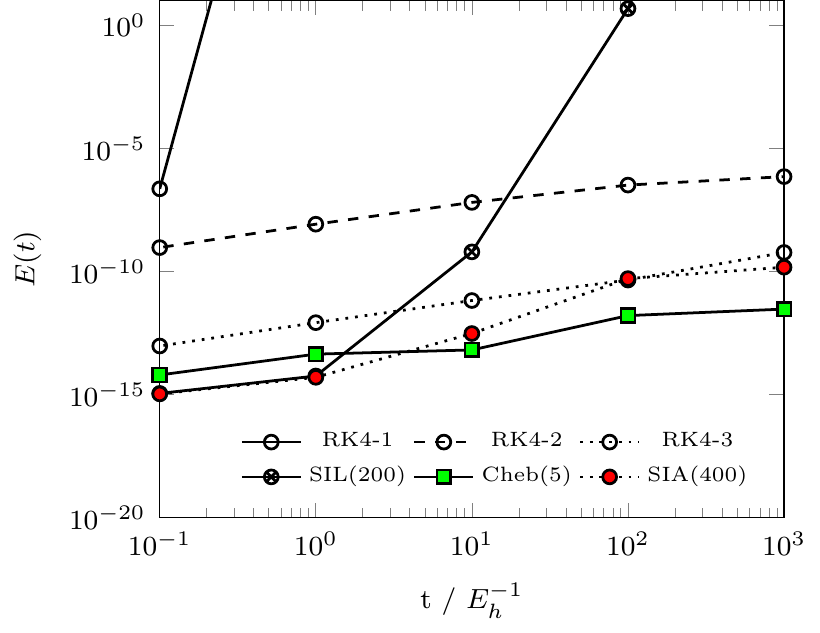}
    \caption{MgF}
  \end{subfigure}
  \caption{Accumulated $S(t)$ errors for RK4, Chebyshev, SIA, and SIL.}
  \label{fig:td_error}
\end{figure}

First, we examine the temporal error accumulation in the autocorrelation
function (\cref{eq:osc_str_func}) using the normalized
root-mean-square-deviation (RMSD) metric
\begin{equation}
  E(t_j) = \sqrt{\frac{\sum_{i\leq j} \vert S(t_i) - S_\mathrm{ex}(t_i)\vert^2}{\sum_{i\leq j} \vert S_\mathrm{ex}(t_i)\vert^2}}, 
  \quad t_i = i\times \delta t,
\end{equation}
where $S_\mathrm{ex}$ is given in \cref{eq:exact_s} and $\delta t$ is
the temporal resolution of the integrated time series.  For the Chebyshev, SIA,
and SIL integrators, $\delta t = 0.05\;E_h^{-1}$.  As the temporal resolution
and step-size coincide for RK4, we have compared our methods with 3 different
RK4 step-sizes to illustrate convergence: 
RK4-1 ($\delta t = 0.05\;E_h^{-1}$),
RK4-2 ($\delta t = 0.01\;E_h^{-1}$), and
RK4-3 ($\delta t = 0.001\;E_h^{-1}$). 
In the following, we will use $E(\mathcal{T})$ (i.e. the total accumulated
autocorrelation error) as a global error metric to assess each integrators'
relative accuracy. \Cref{fig:td_error} illustrates the accumulated
autocorrelation error for each of the integrators considered.  Parameters for
Chebyshev ($\Delta t_\mathrm{cheb}$), SIA ($k$), and SIL ($k$) simulations in
\cref{fig:td_error} were selected to minimize $E(\mathcal{T})$ for each method.
For \Nmol, the Chebyshev, SIA and RK4 integrators exhibit near constant error
accumulation over the full simulation. SIL exhibits a sharp error increase
between 1-10 $E_h^{-1}$ which is of the same order as
$\varepsilon^\mathrm{krylov}$. For $k=36$, SIA yields an invariant subspace up
to an error of $O(\varepsilon^\mathrm{krylov})$, and as such, the entire
simulation ($t < \mathcal{T}$) can be performed using a single Krylov subspace.
For MgF, SIL and RK4-1 diverge, while Chebyshev, SIA, RK4-2 and RK4-3 exhibit
similar error accumulation characteristics as were observed for \Nmol. However,
unlike \Nmol, SIA does not yield an invariant subspace even with largest
subspace of $k=400$, and thus multiple Krylov subspaces must be generated over
the course of the simulation. As such, error $O(\varepsilon^\mathrm{krylov})$
is compounded at each macro-time step, which explains the overtaking of SIA by
Chebyshev in the long-$t$ limit.

\begin{figure}[t]
  \centering
  \begin{subfigure}[b]{0.48\textwidth}
    \centering
    \importlocalfigure{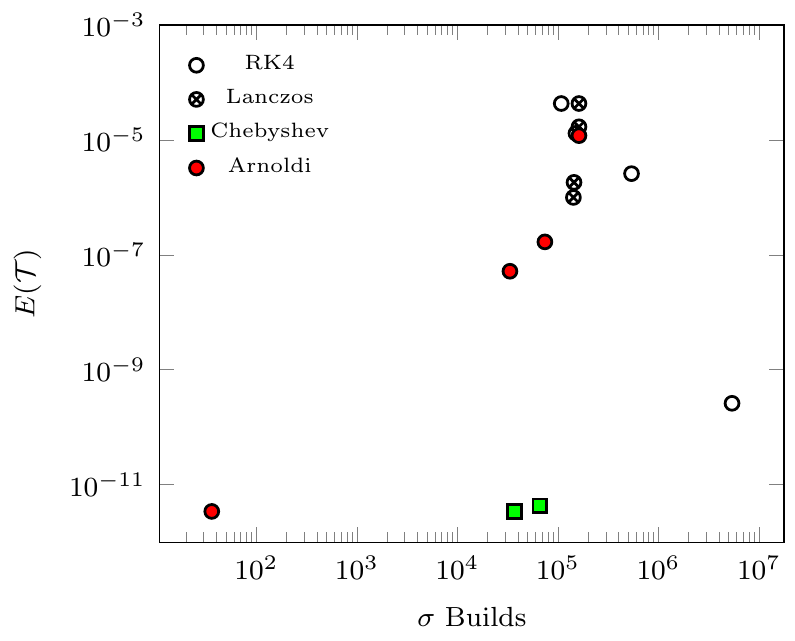}
    \caption{\Nmol}
  \end{subfigure}\hfill%
  \begin{subfigure}[b]{0.48\textwidth}
    \centering
    \importlocalfigure{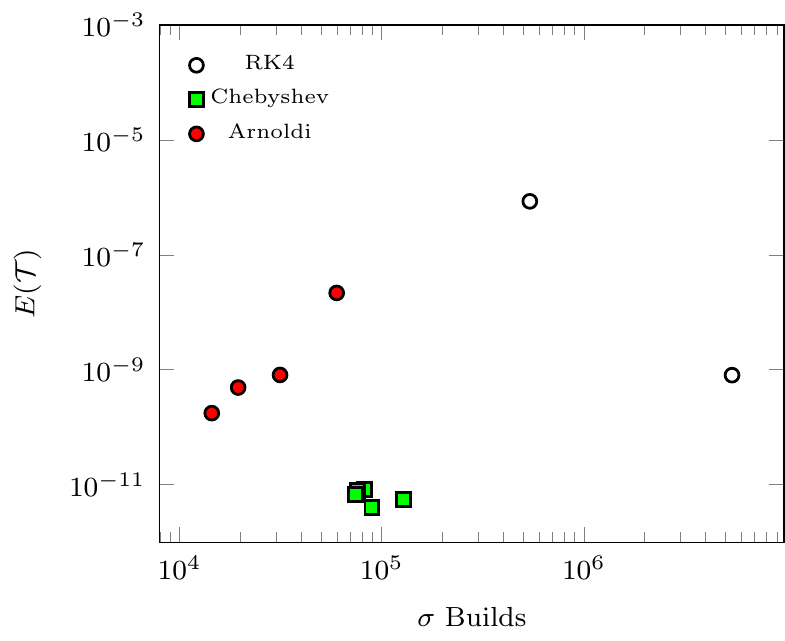}
    \caption{MgF}
    \label{fig:td_cost_mgf}
  \end{subfigure}
  \caption{Cost-to-accuracy comparison for RK4, Chebyshev, SIA, and SIL.}
  \label{fig:td_cost}
\end{figure}

\begin{figure}[t]
  \centering
  \begin{subfigure}[b]{0.45\textwidth}
    \centering
    \importlocalfigure{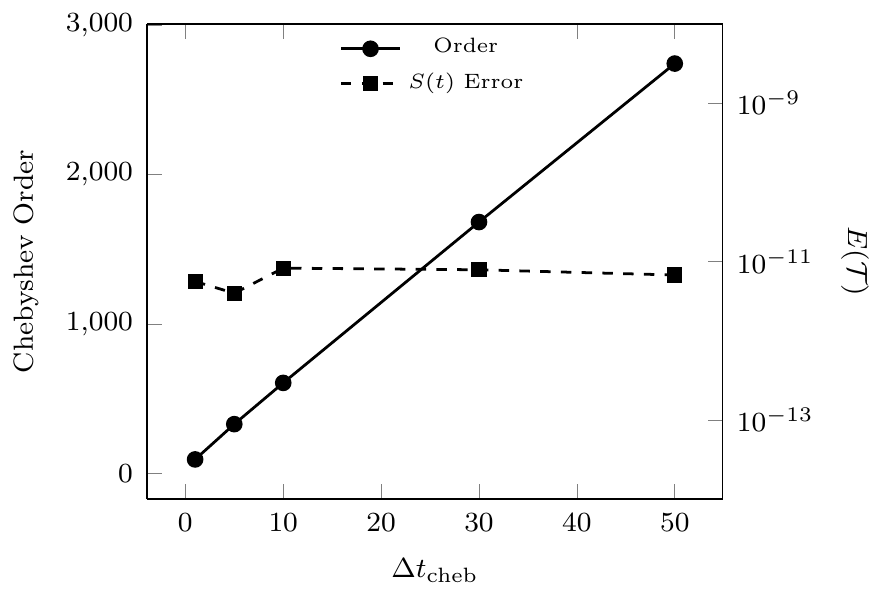}
    \caption{}
    \label{fig:param_scaling_cheb}
  \end{subfigure}\hfill%
  \begin{subfigure}[b]{0.45\textwidth}
    \centering
    \importlocalfigure{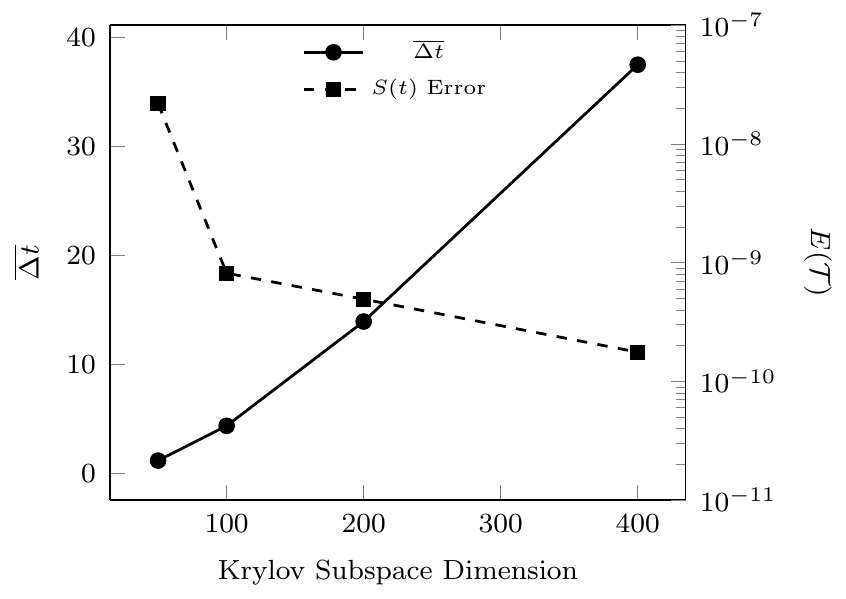}
    \caption{}
    \label{fig:param_scaling_sia}
  \end{subfigure}
  \caption{Assessment of the variance of cost and accuracy of (a) Chebyshev and
  (b) SIA integrators as a function of parameter selection. SIA results are
  presented as the average time-step $\overline{\Delta t}$ as a function of $k$.}
  \label{fig:param_scaling}
\end{figure}

\Cref{fig:td_cost} presents the cost-to-accuracy ratio, characterized by
$E(\mathcal{T})$ as a function of $\sigma$ builds emitted by each integrator,
for a range of parameter choices. For \Nmol~(MgF), Chebyshev results were
obtained for $\Delta t_\mathrm{cheb}\in\{1,5\}$ ($\Delta
t_\mathrm{cheb}\in\{1,5,10,30,50\}$). As discussed in \cref{sec:cheb}, the
number of required $\sigma$ builds for the Chebyshev is fixed at
$m_\mathrm{cheb}\mathcal{T}/\Delta t_\mathrm{cheb}$ and $m_\mathrm{cheb}$
generally increases as a function of $\Delta t_\mathrm{cheb}$. This behaviour
is shown explicitly for MgF in \cref{fig:param_scaling_cheb}.  For both systems
studied, neither $E(\mathcal{T})$ nor to the total number of
$\sigma$-formations are significantly affected by increasing 
$\Delta t_\mathrm{cheb}$.

SIA results were obtained for \Nmol~(MgF) with $k\in\{5,10,20,36\}$
($k\in\{50,100,200,400\}$).  As is shown in \cref{fig:param_scaling_sia}, the
achievable time step ($\sigma$ build count) subject to
$\varepsilon^\mathrm{krylov}$ is (inversely) proportional to $k$ and thus the
SIA and SIL data points in \cref{fig:td_cost} are plotted in order of
\emph{decreasing} $k$.  Unlike the Chebyshev method, the accuracy of SIA
consistently improves with increased $k$, and thus $k$ should be maximized
subject to available memory resources to improve both accuracy and efficiency
of the SIA method. 

For \Nmol, SIL results were also obtained with $k\in\{5,10,20,36,50\}$ for a
direct order-by-order comparison with SIA. At each order, SIA achieves between
2-3 orders of magnitude better accuracy over SIL, and requires $>$50\%
fewer $\sigma$ builds in cases where SIA is able to take time-steps lager than
$\delta t$ ($k \geq 10$). This is due to the fact that the Arnoldi method
generates a faithful Krylov subspace representation $\eomH$  while
the Lanczos method, being only valid for Hermitian matrices, does not.  This
fact is particularly apparent in SIA's generation of an invariant subspace for
$k=36$ while SIL fails to demonstrate similar convergence.

For all problems considered, the proposed SIA and Chebyshev integrators exhibit
superior accuracy and efficiency over analogous SIL and RK4 simulations.  While
it is possible for RK4 to yield reasonable accuracy at small time-steps
(RK4-3), these simulations require excessive number of $\sigma$ builds and
would not be practical for the simulation of realistic TD-EOM-CC problems.

\section{Conclusions}
\label{sec:conclusions}

In this work, we have presented two approximate exponential time-integrators
for TD-EOM-CC theory based on Chebyshev and Arnoldi (SIA) expansions of the
quantum propagator.  The efficacies of these integrators were demonstrated via
comparison with exact exponential dynamics for two small test problems. The
Chebyshev and SIA integrators were demonstrated to yield superior accuracy and
efficiency when compared to RK4 and the recently developed SIL method for
TD-EOM-CC \cite{Cooper21_Short}. As both of the presented methods are built
from standard algorithmic components required for any implementation of
(TD-)EOM-CC, the implementation of these methods has a low barrier for entry
and holds the potential to yield significant performance and accurate
improvements for these simulations in the future.

The practical application of the presented schemes requires consideration of the balance between desired integration accuracy 
and available computational resources. If memory capacity allows, the SIA method
would be preferred for most chemistry applications due to its systematic improvability 
with respect to truncation order. However, the memory requirement of SIA quickly becomes
prohibitive for large problems and the explicit orthogonalization requirement complicated
efficient distributed memory implementations. In these instances, the Chebyshev method
would be preferred due to its low memory requirement and the simplicity of its implementation.

While the results presented in this work have focused on the moment-based
formalism of TD-EOM-CC, the presented efficacy experiments serve as an
important proof-of-concept to demonstrate the the proposed methods for general
TD-EOM-CC simulations. Future work to extend these methods to large scale
TD-EOM-CC simulations is currently being pursued by the authors.  Further,
extension of these methods for use with time-dependent Hamiltonians, such as
those required to study field-driven dynamics of molecular systems, are
currently under development.

\begin{acknowledgement}
This material is based upon work supported by the U.S. Department of Energy,
Office of Science, Office of Advanced Scientific Computing Research and Office
of Basic Energy Sciences, Scientific Discovery through the Advanced Computing
(SciDAC) program under Award No. DE-SC0022263. This project used resources of
the National Energy Research Scientific Computing Center, a DOE Office of
Science User Facility supported by the Office of Science of the U.S. Department
of Energy under Contract No. DE-AC02-05CH11231 using NERSC award ERCAP-0024336.
\end{acknowledgement}

\bibliography{refs}
\end{document}